\documentclass[12pt,number,sort&compress]{elsarticle}

\journal{Nucl. Instr. and Meth. A}

\newcommand{\mathbold}[1]{\mbox{\boldmath $#1$}}
\newcommand{\sla}[1]{\not{\! {#1}}}
\newcommand{\bra}[1]{\left< {#1} \,\right\vert}
\newcommand{\ket}[1]{\left\vert\, {#1} \, \right>}
\newcommand{\braket}[2]{\hbox{$\left< {#1} \,\vrule\, {#2} \right>$}}
\newcommand{\gsim}{\hbox{ \raise3pt\hbox to 0pt{$>$}\raise-3pt\hbox{$\sim$} }}
\newcommand{\lsim}{\hbox{ \raise3pt\hbox to 0pt{$<$}\raise-3pt\hbox{$\sim$} }}
\newcommand{\del}{\ifmmode{\nabla}               \else{$\nabla$ }               \fi}

\newcommand{\figdir}{.}

\begin{document}

\begin{frontmatter}



\title {An estimation of the effective number of electrons \\
           contributing to the coordinate measurement with a TPC \\
                             II}

%
\author{Makoto Kobayashi}
\address{High Energy Accelerator Research Organization (KEK),
                                         Tsukuba, 305-0801, Japan}
%
%
%
%
%
%


\begin{abstract}
For time projection chambers (TPCs) the accuracy in measurement of 
the track coordinates along the pad-row direction deteriorates with the
drift distance ($z$):
$\sigma_{\rm X}^2 \sim D^2 \cdot z / N_{\rm eff}$, where
$D$ is the diffusion constant and $N_{\rm eff}$ is the
effective number of electrons.
Experimentally it has been shown that $N_{\rm eff}$ is smaller than
the average number of drift electrons per pad row ($\left< N \right>$).
In the previous work we estimated  $N_{\rm eff}$ by means of a
simple numerical simulation for argon-based gas mixtures, taking into account the
diffusion of electrons only in the pad-row direction \cite{Kobayashi1}.  
The simulation showed that $N_{\rm eff}$ could be as small as $\sim$ 30\% 
of $\left< N \right>$ because of the combined effect of statistical
fluctuations in the number of drift electrons ($N$) and in their
multiplication in avalanches.
In this paper, we evaluate the influence of the diffusion normal to the
pad-row direction on the effective number of electrons.
The de-clustering of the drift electrons due to the diffusion makes $N_{\rm eff}$ 
drift-distance dependent. However, its effect was found to be too small
to explain the discrepancy between the values of $N_{\rm eff}$
measured with two TPC prototypes different in size.

\end{abstract}
%
%
%
%

\end{frontmatter}


%
%
%
%
%
%

\section{Introduction}

In the previous paper we estimated the effective number of electrons
($N_{\rm eff}$) contributing to the coordinate measurement of a 
time projection chamber (TPC) equipped with a Micro-Pattern Gaseous Detector
(MPGD) and ideal readout electronics \cite{Kobayashi1}.
$N_{\rm eff}$ parametrizes the spatial resolution for a pad row as follows
\begin{equation}
\sigma_{\rm X}^2 = \sigma_{\rm X0}^2 + \frac{D^2}{N_{\rm eff}} \cdot z \label{eq1}
\end{equation}
where $\sigma_{\rm X}$ is the spatial resolution along the
pad-row direction, $\sigma_{\rm X0}$ is the intrinsic resolution, and
$D$ denotes the transverse diffusion constant, with $z$ being the drift
distance.
In the ideal case (with an infinitesimal pad pitch) $\sigma_{\rm X0}$
vanishes for particle tracks perpendicular to the pad row, and increases
with the track angle ($\phi$) with respect to the pad-row normal
because of the angular pad effect.
Only right angle tracks ($\phi$ = 0$^\circ$) are considered throughout
the present work.
 
Under the conditions listed in Ref.~\cite{Kobayashi1}, $N_{\rm eff}$  is given by
\begin{equation}
\frac{1}{N_{\rm eff}} = \left< \frac{1}{N} \right> \cdot ( 1 + f ) \label{eq2}
\end{equation}
where $N$ denotes the total number of drift electrons detected by the pad row
and $f$ is the relative variance of the gas-amplified signal charge
($q$) induced on the pad row by single drift electrons
(avalanche fluctuation: $\sigma_q^2 / \left<q\right>^2$)\footnote{
It should be noted that Eq.~(\ref{eq2}) is an expression for $1/N(h)$ in
Eq.~(7.33) of Ref.~\cite{Blum}.}.
Although Eq.~(\ref{eq2}) was derived assuming the total charge 
($\sum_{i=1}^N q_i$) to be constant ($N \cdot \left< q \right>$)
it was found to be a good approximation by a numerical simulation for a
practical value of $\left< N \right>$ (see Ref.~\cite{Kobayashi1},
and Appendix A for details).

The simulation for argon-based gas mixtures gave $N_{\rm eff}$ of $\sim$
22 for $f = 2/3$ and a pad-row pitch of 6.3 mm \cite{Kobayashi1},
which is about 30\% of the average value of $N$ ($\left< N \right> \sim 71$),
and is consistent with the values obtained with a small prototype TPC
\cite{Kobayashi2, Paul, Kobayashi3}.
The value of $N_{\rm eff}$ corresponds to $\sim 36$ for a pad-row
pitch normalized to 1 cm, assuming $N_{\rm eff}$ to be (approximately) proportional
to the pad-row pitch\footnote{
Actually, this is a bold assumption. See Appendix A.}.

Recent resolution measurements with a larger prototype TPC with
MicroMEGAS readout, however, gave a significantly larger estimate
for $N_{\rm eff}$ ($\sim 56$ for 1-cm pad height) \cite{Paul2}\footnote{
In fact, the values of  $N_{\rm eff}$ were obtained using
different kinds of charged particle: a beam of 5-GeV/$c$ electrons in Ref.~\cite{Paul2}
while a beam of 4-GeV/$c$ pions or cosmic rays in 
Refs.~\cite{Kobayashi2, Paul, Kobayashi3}.
The discrepancy is still large, however, even if the difference in the
primary ionization density is taken into account.
}.
A possible origin of the discrepancy could be the de-clustering of drift
electrons due to diffusion normal to the pad-row direction ($D_{\rm y}$),
which is expected to be more efficient for larger TPCs with a longer
average drift distance.
It should be pointed out that the diffusion only along the pad-row
direction ($D_{\rm x}$) was taken into account in Ref.~\cite{Kobayashi1}.
In the present work, we evaluate the contribution of the
de-clustering effect to the increase of $N_{\rm eff}$, through the
decrease of $\left< 1/N \right>$ due to the finite $D_{\rm y}$,
in argon-based gas mixtures.

An analytic and qualitative approach is described through
Section 2 to 4,
the results of a numerical simulation are shown in Section 5,
and Section 6 concludes the paper.
Readers are suggested to read Ref.~\cite{Kobayashi1} in advance.

\section{Long drift-distance limit}

Let us consider the hypothetical case of a TPC with an infinitely large
drift distance\footnote{
The dimensions of the readout pad plane are considered to be
infinitely large as well.
Otherwise a part of the drift electrons created at long drift distances
would be absorbed by the field cage (the inner or outer wall of
a cylindrical TPC) before reaching the readout plane.}.
Primary ionization clusters created along a particle track
at an infinitely large drift distance get completely de-clustered,
and the secondary electrons distribute uniformly and randomly on the readout plane.
They no longer have any information on the original cluster positions. 
Therefore the total number of drift electrons reaching a pad row obeys
Poisson statistics with a mean $\mu = \left< N \right>$:
\begin{displaymath} 
  P(N) = e^{-\mu} \cdot \frac{\mu^N}{N!} \; .  
\end{displaymath}

\begin{figure}[htbp]
\begin{center}
\hspace{10mm}
\includegraphics*[scale=0.55]{\figdir/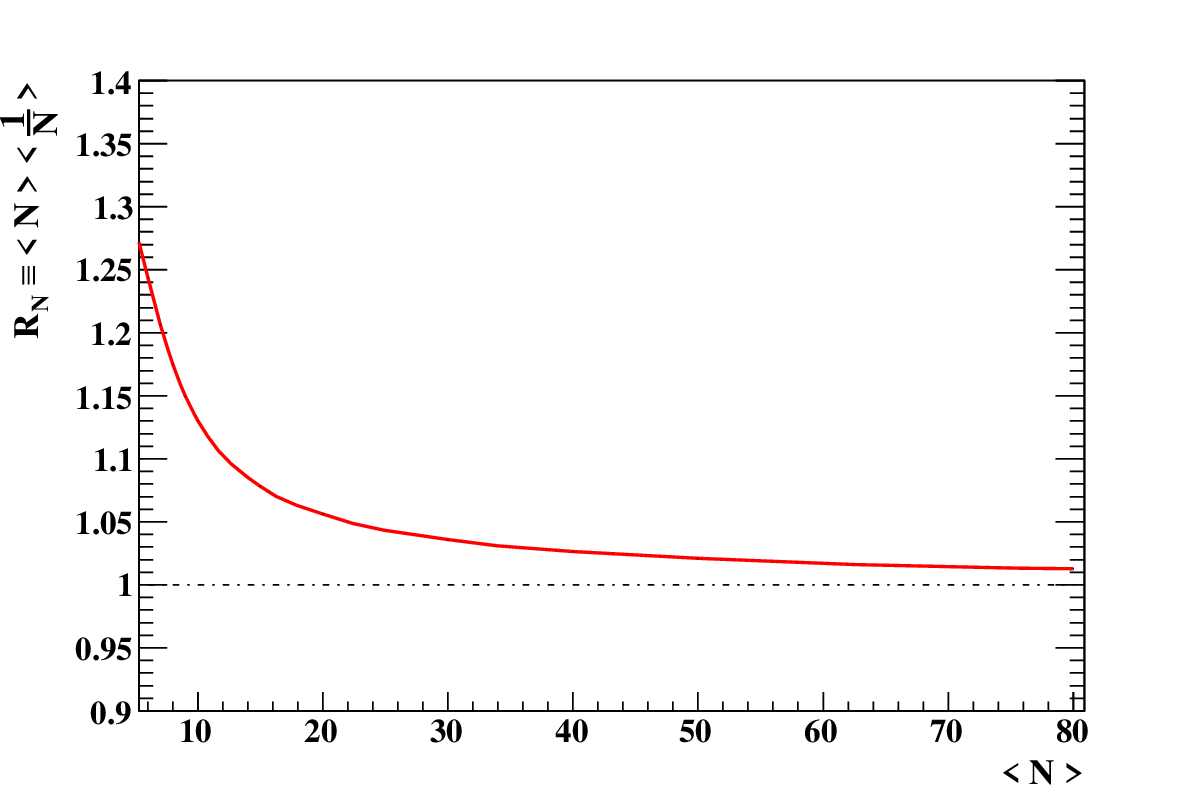}
\end{center}
\caption{\label{fig1}
\footnotesize $R_{\rm N} \equiv \left< N \right> \cdot
\left< 1/N \right>$ as a function of the average number of electrons
in the case of Poisson distribution for $N$. 
}
\end{figure}

The average value of the inverse of $N$ in this case is given by 
\begin{eqnarray*}
\left< \frac{1}{N} \right> &=& \frac{1}{1 - e^{-\mu}} \cdot 
                              \sum_{N=1}^\infty \frac{1}{N} \cdot P(N) \\
&=& \frac{e^{-\mu}}{1 - e^{-\mu}} \cdot \sum_{N=1}^\infty \frac{\mu^N}{N \cdot N!} \\
&=&  \frac{e^{-\mu}}{1 - e^{-\mu}} \cdot \left( E_i(\mu) - \ln (\mu) - \gamma \right)
\end{eqnarray*}
where $E_i$ is the exponential integral\footnote{
The exponential integral is defined as
\[
E_i(x)
 = \int_{-\infty}^x \frac{e^t}{t} \; dt \; .
\]
}
and $\gamma$ is the Euler-Mascheroni constant ($\sim$ 0.577).
When $N = 0$ the pad row is inefficient and provides no coordinate
measurement.
Therefore it is excluded from the summation.  
Fig.~\ref{fig1} shows the behavior of $R_{\rm N} \equiv \left< N \right> \cdot
\left< 1/N \right>$ as a function of $\left< N \right>$.
For a practical value of $\left< N \right> \gsim 50$,
$\left< 1/N \right> \sim 1/ \left< N \right>$ and
$N_{\rm eff}$ is expected to be $\sim \left< N \right> / (1 + f)$
at an infinitely long drift distance.

\section{Short drift-distance limit}

At zero drift distance, where the clusters are intact, 
the $N$-distribution is a Landau type with a
long tail for large $N$ due to occasional large clusters.
For the Landau distribution, the mode ($\hat{N}$) is considerably
smaller than the average ($\left< N \right>$), whereas
$\left< 1/N \right >$ is close to 1/$\hat{N}$
(see, for example, Fig.~5 and 6 in Ref.~\cite{Kobayashi1}).
Therefore $R_{\rm N}$ defined above is significantly larger than unity even for
relatively large $\left< N \right>$ and 
$N_{\rm eff} = \left< N \right> / (1 + f) / R_{\rm N} 
 < \left< N \right> / (1 + f)$.    

As the clusters disintegrate with the increase of drift distance,
the $N$-distribution changes its shape because of the diffusion normal
to the pad-row direction, approaching a 
Poissonian, for which $\hat{N} \sim \left< N \right>$\footnote{
A Poissonian with $\left< N \right> \gsim 20$ is close to a Gaussian.
}. 
With the progress of de-clustering $\hat{N}$ shifts towards $\left< N \right>$,
therefore $\left< 1/N \right >$ decreases, while
$\left< N \right>$ remains constant\footnote{
The most probable energy loss measured with a pad row is therefore 
expected to increase gradually with the drift distance.
}.
Consequently, $N_{\rm eff}$ is expected to be an increasing function of the
drift distance, with an asymptotic maximum of 
$\sim \left< N \right>/(1 + f)$ .

The rate at which the Landau distribution approaches a Poissonian
with the increase of drift distance depends on the pad height,
the diffusion constant, and the cluster size distribution.
In the next section, the change in the variance of the
$N$-distribution is calculated in order to demonstrate that
the transition to the Poissonian is slow.   

\section{Variance of the $N$-distribution}

We evaluate in this section the variance (Var) of the total number of electrons
reaching a readout pad row ($N$) since Var($N$) gives a good measure for
the deviation of the $N$-distribution from a Poissonian, for which
Var\/($N$) = $\left< N \right>$.
First, let us suppose that a single point-like electron cluster of size
$n$ is created at a coordinate $y = Y$ in the direction of the pad-row
normal and $z = Z$ in the drift direction, measured from the readout
plane.
The electrons diffuse on their way towards the readout plane.
Their spread in the y-coordinate is given by a
Gaussian with a standard deviation of
$\sigma \equiv \sigma_y = D \cdot\sqrt{Z}$,
at the pad rows with a height of $h$\footnote{
More precisely $h$ should be understood as the pad-row pitch,
which is usually slightly larger than the pad height
when the readout plane is covered over with pads.
The pad-row pitch and the pad height ($h$) are not distinguished in the 
present paper.
}.
The probability to find $\nu$ electrons reaching the pad row is
given by\footnote{
Hereafter the notation $f^n(x)$ represents the $n$-th power of a function
$f(x)$, i.e. $(f(x))^n$.
} 
\begin{displaymath}
P(\nu ; Y,\sigma) = {{n}\choose{\nu}} \cdot \Pi^\nu(Y) 
      \cdot \left( 1 - \Pi(Y) \right)^{n - \nu}
\end{displaymath}
where
\begin{displaymath}
\Pi(Y) = \int_{-h/2}^{h/2} G(y;Y,\sigma) \; dy
\end{displaymath}
with
\begin{displaymath}
G(y;Y,\sigma) \equiv \frac{1}{\sqrt{2\pi} \sigma} \cdot 
          \exp \left[ - \frac{(y-Y)^2}{2\sigma^2} \right] \; .
\end{displaymath}
Since $P(\nu ; Y,\sigma)$ represents a binomial statistics for a fixed
$Y$ (and $\sigma$), $\left< \nu \right>$ and $\left< \nu^2 \right>$ are given by
\begin{displaymath}
\left< \nu(Y) \right> = n \cdot \Pi(Y)
\end{displaymath}
and
\begin{displaymath}
\left< \nu^2(Y) \right> = n \cdot \Pi(Y)
     + n(n-1) \cdot \Pi^2(Y) \; .
\end{displaymath}

\begin{figure}[htbp]
\begin{center}
\hspace{10mm}
\includegraphics*[scale=0.46]{\figdir/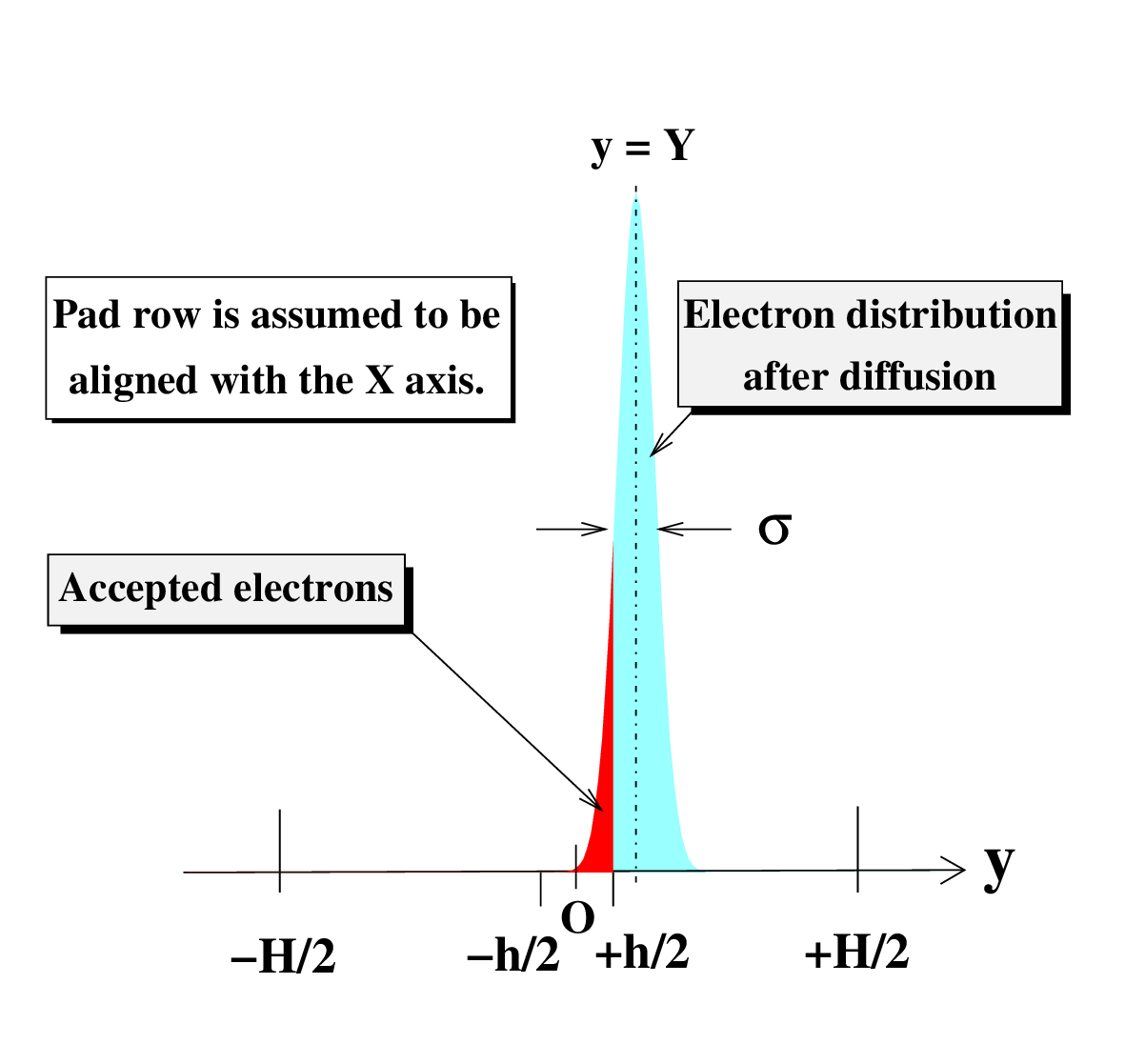}
\end{center}
\caption{\label{fig2}
\footnotesize Illustration of the relevant variables. 
}
\end{figure}

Let us further assume that the initial cluster is randomly created in a
$y$-region [$-H/2, +H/2$] with $H \gg \sigma$ (see Fig.~\ref{fig2}).
Then, averaging over $-H/2 \leq Y \leq +H/2$, 
\begin{eqnarray*}
\left< \nu \right> &=& \frac{1}{H} \int_{-H/2}^{H/2} 
                       \left< \nu(Y) \right> dY \\
&=& \frac{n}{H} \int_{-H/2}^{H/2} dY
         \int_{-h/2}^{h/2} G(y;Y,\sigma) \; dy \\
&=& \frac{n}{H} \int_{-h/2}^{h/2} dy
         \int_{-H/2}^{H/2} G(y;Y,\sigma) \; dY \\
&\sim& \frac{n \cdot h}{H} \\
{\rm Var}(\nu) &=& \frac{1}{H} \int_{-H/2}^{H/2} 
            \left< \nu^2(Y) \right> \; dy 
                    - \left< \nu \right>^2  \\
&\sim& \frac{n \cdot h}{H}
  + \frac{n(n-1)}{H}\int_{-H/2}^{H/2} \Pi^2(Y) \; dY
       - \left( \frac{n \cdot h}{H} \right)^2 \\
&\sim&  \frac{n \cdot h}{H}
  + \frac{n(n-1)}{H} \cdot h \cdot g\left( \frac{\sigma}{h} \right)
       - \left( \frac{n \cdot h}{H} \right)^2
\end{eqnarray*}
where 
\begin{displaymath}
g\left( \frac{\sigma}{h} \right) \equiv
   {\rm erf}\left( \frac{h}{2\sigma} \right)
  - \frac{2}{\sqrt{\pi}} \cdot \frac{\sigma}{h} \cdot
  \left[ 1 - \exp \left( - \frac{h^2}{4\sigma^2} \right) \right]  \; .
\end{displaymath}
See Appendix B for the derivation of the function $g(\sigma/h)$. 

If there are $N_{\rm CL}$ (independent) clusters in $[-H/2,+H/2]$,
the average and the variance are given by multiplying
$N_{\rm CL}$\footnote{
Note that
\begin{displaymath}
\left< \sum_{i=1}^{N_{\rm CL}}\nu_i \right> =
  N_{\rm CL} \cdot \left< \nu \right> \hspace{3mm} {\rm and} \hspace{71mm}
\end{displaymath}
\begin{displaymath}
\left< \left( \sum_{i=1}^{N_{\rm CL}} \nu_i
   - \left< \sum_{i=1}^{N_{\rm CL}} \nu_i \right> \right)^2 \right>
 = \left< \left( \sum_{i=1}^{N_{\rm CL}} \left( \nu_i -
	            \left< \nu_i\right> \right) \right)^2 \right>
 = N_{\rm CL} \cdot 
      \left< \left( \nu - \left< \nu \right> \right)^2 \right> \; .
\end{displaymath}
}:
\begin{eqnarray*}
\left< \nu \right> &=& \frac{N_{\rm CL}}{H} \cdot n \cdot h \\
{\rm Var}(\nu) &=& \frac{N_{\rm CL}}{H} \cdot n \cdot h
 + \frac{N_{\rm CL}}{H} \cdot n(n-1) \cdot h \cdot g\left( \frac{\sigma}{h} \right) 
 - N_{\rm CL} \cdot \left( \frac{n \cdot h}{H} \right)^2 \; .
\end{eqnarray*}
Taking the limit of $H \rightarrow \infty$ while keeping the cluster
density $\rho \equiv N_{\rm CL}/H$ constant,
\begin{eqnarray*}
\left< \nu \right> &=& \rho \cdot n \cdot h \\
{\rm Var}(\nu) &=&  \rho \cdot n \cdot h + \rho \cdot n(n-1) \cdot h \cdot
                   g\left( \frac{\sigma}{h} \right) \; .  
\end{eqnarray*}

In reality the cluster density ($\rho$) depends on the cluster size
($n$):
\begin{displaymath}
 \rho = \rho(n) \equiv \rho_0 \cdot p\/(n)
\end{displaymath}
where $\rho_0$ is the total primary ionization density and $p\/(n)$
is the proportion of the cluster of size $n$
($\sum_n p\/(n) = 1$).
The number of electrons detected by the pad row ($N$) is the sum of
the contribution ($\nu$) from various cluster sizes.
Consequently, its average and variance are given by
\begin{eqnarray*}
\left< N \right> &=& \rho_0 \cdot h \cdot \sum_n n \cdot p\/(n) \\
{\rm Var}(N) &=& \rho_0 \cdot h \cdot \sum_{n} n \cdot p\/(n) \cdot
     \left( 1 + (n-1) \cdot g \left( \frac{\sigma}{h}  \right)  \right)
     \; .
\end{eqnarray*}

\begin{figure}[htbp]
\begin{center}
\hspace{10mm}
\includegraphics*[scale=0.60]{\figdir/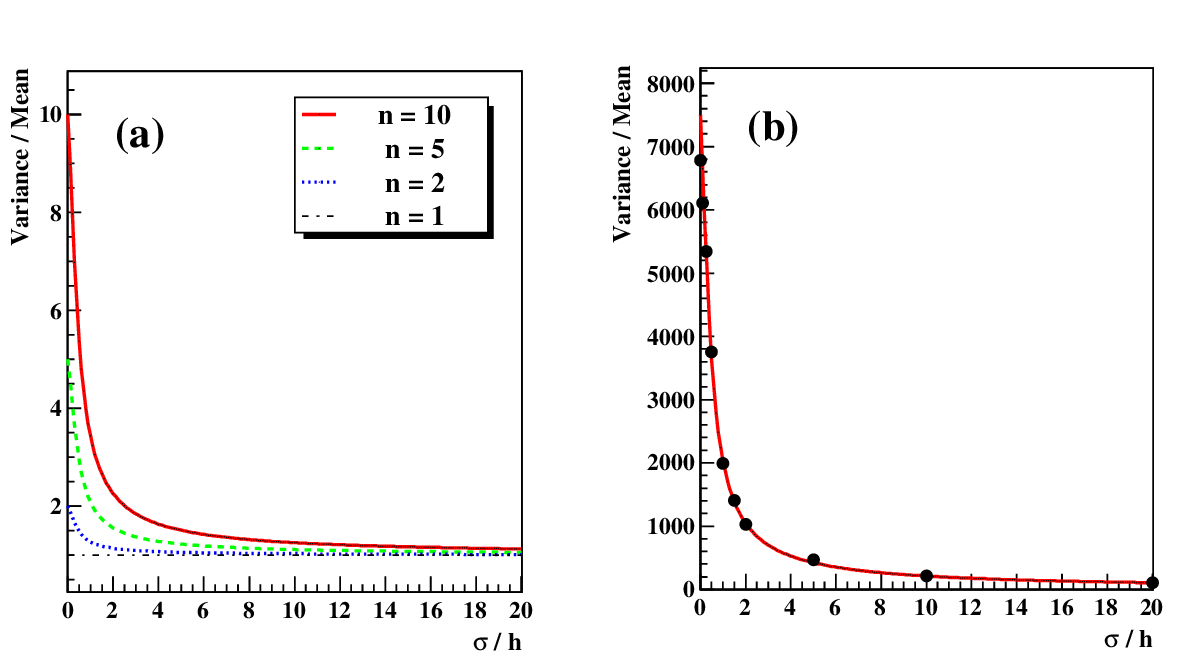}
\end{center}
\caption{\label{fig3}
\footnotesize Variance divided by mean as a function of the scaling
parameter $\sigma / h$: (a) for several fixed values of the cluster size
and (b) for a realistic cluster-size distribution along with the 
{\it data points\/} given by a numerical simulation (filled circles).  
 }
\end{figure}

Fig.~\ref{fig3}~(a) shows
\begin{displaymath}
\frac{{\rm Var}(\nu)}{\left< \nu \right>} = 
    1 + (n-1) \cdot g \left( \frac{\sigma}{h}  \right) 
\end{displaymath}
for several values of the cluster size $n$
as a function of the scaling parameter $\sigma /h $.
It is clear that the $\nu$ distribution approaches a Poissonian
(${\rm Var}(\nu) / \left< \nu \right> = 1$) slowly with the increase of
$\sigma / h$, especially for large clusters.
Fig.~\ref{fig3}~(b) shows the variance divided by the average for a realistic 
$N$-distribution:
\begin{eqnarray*}
\frac{{\rm Var}(N)}{\left< N \right>} &=&
   \frac
     {\sum_n n \cdot p\/(n)
        \cdot \left( 1 + 
     (n-1) \cdot g \left( \frac{\sigma}{h} \right) \right)}
  {\sum_n n \cdot p\/(n)} \\
&=& 1 +
   \frac
     {\sum_n n(n-1) \cdot p\/(n)}{\sum_n n \cdot p\/(n)}
      \cdot g \left( \frac{\sigma}{h} \right) \; ,
\end{eqnarray*}
along with the ratios obtained with a numerical simulation (see Section 5).
The probability mass function $p\/(n)$ was assumed to be that
corresponding to the cluster-size distribution shown in Fig.~2
of Ref.~\cite{Kobayashi1}.
It should be noted that
${\rm Var}(N)/\left< N \right > = \left< n^2 \right> / \left< n \right>$
whereas  
${\rm Var}(\nu)/\left< \nu \right > = n\; ,$
at $\sigma/h = 0$.
Consequently, the value of ${\rm Var}(N)/\left< N \right >$ at zero drift distance 
is rather large because of the contribution of (very) large clusters.

Fig.~\ref{fig3}~(b) tells us that the $N$-distribution is a Landau type
at short drift distances and approaches a Poissonian very slowly
with the increase of drift distance, owing to the de-clustering.
Its average  ($\left< N \right>$) remains constant during the
transition.

\section{Evaluation of $N_{\rm eff}$ by a simulation}

The analytic approach described through Section 2 to 4 shows that $N_{\rm eff}$
is expected to be a slowly increasing function of $\sigma / h$, i.e. the
drift distance.
In order to confirm this quantitatively, we evaluated $N_{\rm eff}$ by
means of a numerical simulation.

The simulation code is identical to that used in the previous 
work \cite{Kobayashi1}, except that the diffusion of drift electrons
normal to the pad-row direction ($D_{\rm y}$) is taken into account.
Initial electron clusters are randomly generated along the $y$-axis
(with the pad row aligned with the $x$-axis) 
in a range wide enough compared to the
diffusion ($\sigma = D \cdot \sqrt{Z}$) and the pad height ($h$).
The cluster density is assumed to be 24.3 cm$^{-1}$ $\times$
1.2 (relativistic rise factor) as in the previous paper~\cite{Kobayashi1}.
The size of each cluster is determined randomly using the
probability mass function $p\/(n)$ (see Section 4).
The secondary electrons originated from each cluster are then dispersed
in the directions of the pad row ($x$) and the pad-row normal ($y$)
with $\sigma_x = \sigma_y = \sigma = D \cdot \sqrt{Z}$ \footnote{
Note that $D_x = D_y = D$
since the magnetic field (if it exists) is parallel to the $z$-axis.
}. 
The electrons with the final position located within the pad row
($|y| \leq h/2$) are accepted (see Fig~\ref{fig2}).
Gas gain is assigned to each of the accepted electrons randomly assuming 
a Polya distribution ($\theta = 0.5$, corresponding to $f = 2/3$) for the
avalanche fluctuation.   
The coordinate resolution ($\sigma_{\rm X}$) is evaluated from the 
fluctuation in the charge centroid of the accepted electrons in the pad-row
direction ($x$\/).
The square of the ratio of the diffusion ($\sigma$) to the resolution
($\sigma_{\rm X}$) gives $N_{\rm eff}$
from Eq.~(\ref{eq1}) with $\sigma_{\rm X0}$ = 0.

\begin{figure}[htbp]
\begin{center}
\hspace{10mm}
\includegraphics*[scale=0.80]{\figdir/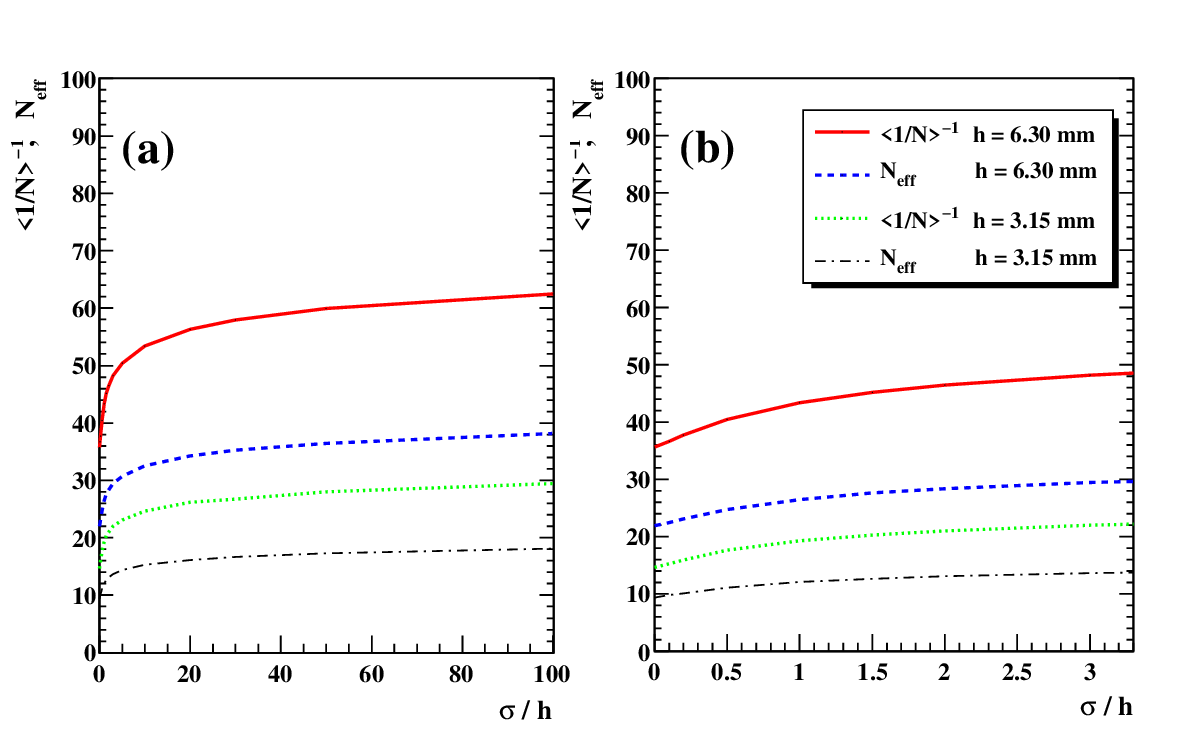}
\end{center}
\caption{\label{fig4}
\footnotesize Inverse of the average of $1/N$ and effective number of
electrons as a function of $\sigma / h$, given by the simulation
for $h$ = 6.30 mm and 3.15 mm.
The values for small (realistic) $\sigma / h$ are shown in (b).
The relative variance of the avalanche fluctuation ($f$) is
taken to be 2/3. The legends are identical for (a) and (b).
}
\end{figure}

Fig.~\ref{fig4} shows the obtained $N_{\rm eff}$ and $\left< 1/N \right>^{-1}$
as a function of $\sigma/h$.
The effective number of electrons certainly increases with $\sigma/h$
in association with the increase of $\left< 1/N \right>^{-1}$.
The asymptotic value of $\left< 1/N \right>^{-1}$ is about 70 (35)
for $h$ = 6.3 mm (3.15 mm) as expected. 
However, the increase of $N_{\rm eff}$ is rather slow and would be observable
only for large values of $\sigma/h$, i.e. at (very) long drift distances.

\begin{figure}[htbp]
\begin{center}
\hspace{10mm}
\includegraphics*[scale=0.55]{\figdir/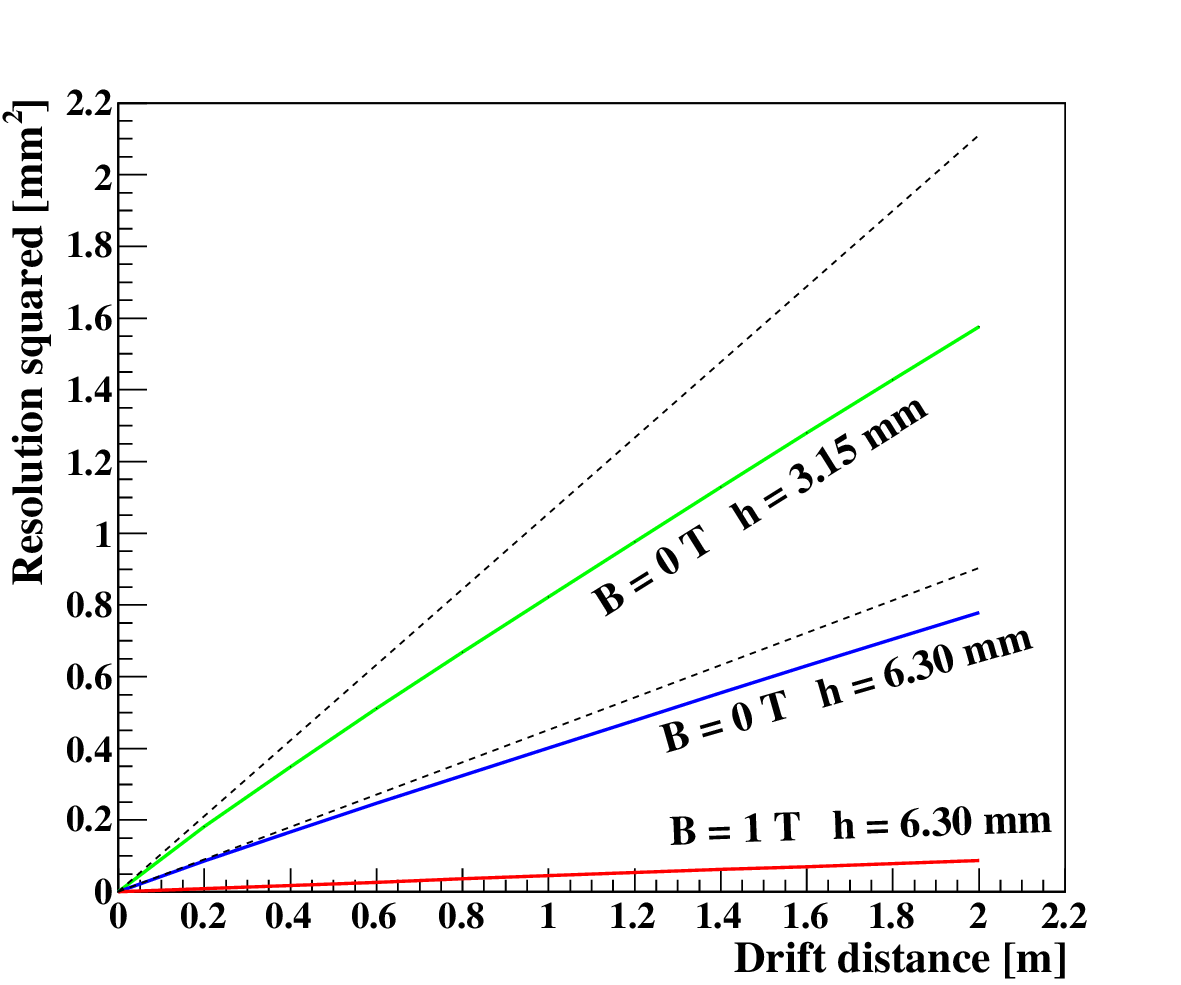}
\end{center}
\caption{\label{fig5}
\footnotesize Resolution squared per pad row as a function of the
drift distance ($z$)
given by the simulation for $B$ = 0 and 1 T, with a pad height ($h$) of 6.3 mm.
The diffusion constant is assumed to be 315 (101) $\mu$m/$\sqrt{\rm cm}$
for $B$= 0 T (1 T) given by Magboltz~\cite{Biagi} for a gas mixture of
Ar-CF4~(3\%)-isobutane~(2\%) and a drift field of 250 V/cm.
The relation simulated for $B$ = 0 T with $h$ = 3.15 mm is also shown
for comparison.
The dashed straight lines show the linear relations with the values of 
$N_{\rm eff}$ fixed to those at $z$ = 0, in the case of $B$ = 0 T.
For $B$ = 1 T, $N_{\rm eff}$ is almost constant throughout the drift
distance shown in the figure.
}
\end{figure}

Examples of the resolution squared as a function of the drift distance
are shown in Fig.~\ref{fig5} for pad heights of 6.3 mm and 3.15 mm. 
The chamber gas is taken to be Ar-CF4 (3\%)-isobutane (2\%) as in the
experiments of Refs.~\cite{Kobayashi3,Paul2}.
The deviation from the linear dependence 
(Eq.~(\ref{eq1}) with a constant $N_{\rm eff}$ evaluated at $z$ = 0) is
prominent at long drift distances without the magnetic field,
in particular for the shorter pad height.

\section{Conclusion}

We estimated the effect of the electron diffusion normal to the
pad-row direction ($D_{\rm y}$) on the spatial resolution of TPCs
operated in argon-based gas mixtures.
It does affect the effective number of electrons 
contributing to the azimuthal coordinate measurement
and thus makes $N_{\rm eff}$ drift-distance dependent:
$N_{\rm eff} = N_{\rm eff}(\sigma/h) = N_{\rm eff}(z)$
for a fixed value of $h$.
The value of $N_{\rm eff}$ increases with the drift distance
since the distribution of the number of electrons detected by a 
pad row asymptotically approaches a Poissonian by de-clustering.
However, the de-clustering process is rather slow
because its scaling parameter is $\sigma/h$,
and $N_{\rm eff}$ can be assumed to be constant for practical TPCs
operated under a strong axial magnetic field.

In addition, the influence of avalanche fluctuation ($R_{\rm q}$) was
found to be almost constant ($\sim 1 + f$) for a realistic pad-row pitch
greater than $\sim$ 6 mm (see Appendix A).
Therefore Eq.~(\ref{eq2}) is expected to give a good approximation
for the value of $N_{\rm eff}$, with $\left< 1/N \right>$ estimated
assuming $D_{\rm y} = 0$.

It is unlikely that the large value of $N_{\rm eff}$ observed with the
larger TPC prototype, with a maximum drift length of $\sim$ 600 mm
and a pad-row pitch of 7 mm,
arises from the finite $D_{\rm y}$.
The larger $N_{\rm eff}$ may have been owing to other factors such as
smaller avalanche fluctuation ($f$), or improvement of the
signal-to-noise ratio and/or
better calibration of the readout electronics
(see Appendix C of Ref.~\cite{Kobayashi3}).
It should be noted that gas contaminants such as oxygen could affect
the apparent value of $N_{\rm eff}$ as well,
through the capture of electrons during their drift towards the
readout plane.

The increase of $N_{\rm eff}$ would be observed at long drift distances
with a large TPC operated in a gas with a relatively large
transverse diffusion constant in the absence of magnetic field
(see Fig.~\ref{fig5}).


\setcounter{figure}{0}

\section*{\nonumber Acknowledgments}
We would like to thank many colleagues of the LCTPC collaboration for their
continuous encouragement and support, and for fruitful discussions.
This work was partly supported by the Specially Promoted Research
Grant No. 23000002 of Japan Society for the Promotion of Science.  

\appendix

\section{Pad-height dependence of $N_{\rm eff}$}
In the previous work, the pad-row pitch ($\sim$ pad height $h$) was fixed to
6.3 mm \cite{Kobayashi1}.
We show here the pad-height dependence of the effective number of electrons
given by a numerical simulation.
The simulation code is exactly the same as that developed for Ref.~\cite{Kobayashi1}.
Therefore the diffusion of drift electrons normal to the pad-row direction
($D_{\rm y}$) is not taken into account and the values of
$N_{\rm eff}$ are those for zero drift distance. 

\begin{figure}[htbp]
\begin{center}
\hspace{10mm}
\includegraphics*[scale=0.55]{\figdir/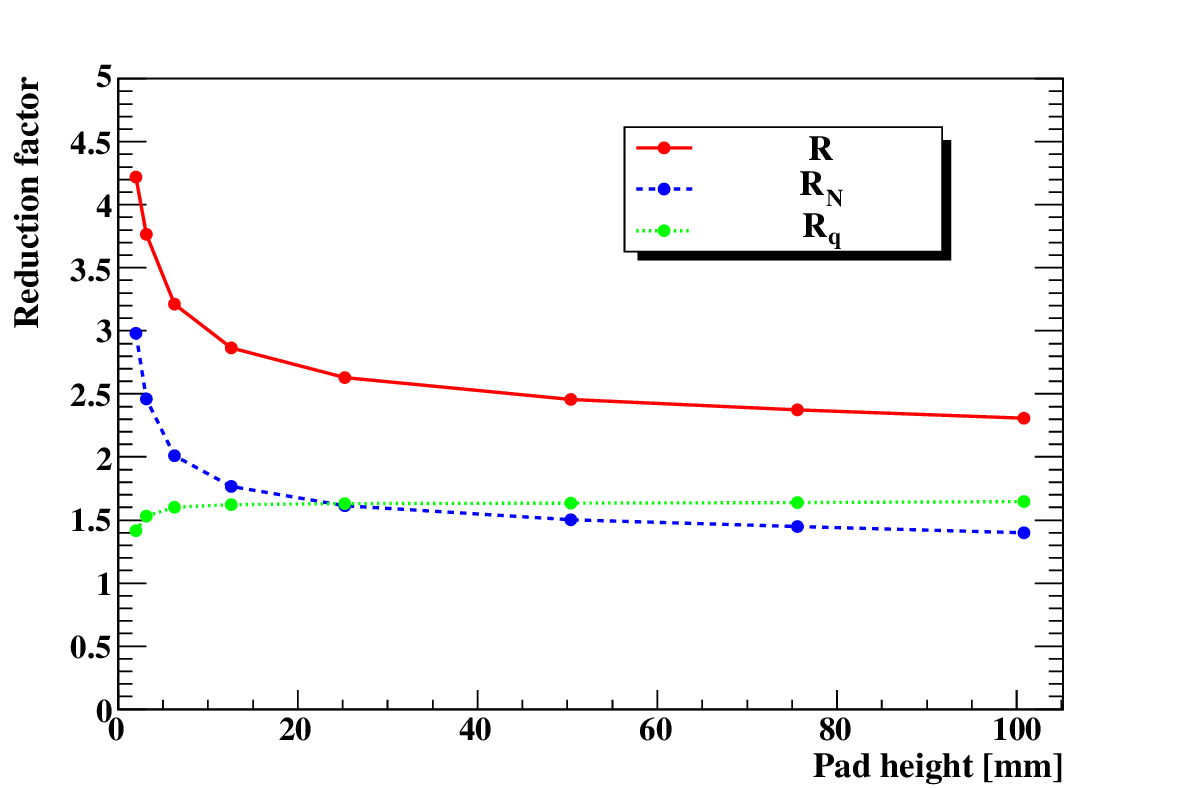}
\end{center}
\caption{\label{figA1}
\footnotesize Reduction factors ($R$, $R_N$ and $R_q$) as a function 
of the pad height.
The relative variance of avalanche fluctuation ($f\/$) is taken to be 2/3.
}
\end{figure}

If we write $N_{\rm eff} = \left< N \right > / R$, with $R$ being a reduction factor,  
$R$ is expressed as $R = R_{\rm N} \cdot R_{\rm q}$, where
$R_{\rm N} = \left< N \right> \cdot \left< 1/N \right>$ and
$R_{\rm q}$ derives from the avalanche fluctuation in the detection
device for each of the drift electrons \cite{Kobayashi1}.
The value of $R_{\rm q}$ is expected to be close to $1 + f$
for large $\left< N \right>$ (large pad height) since
$\sum_{i=1}^N q_i \sim N \cdot \left< q \right>$ becomes a good approximation.
Fig.~\ref{figA1} shows the pad-height dependences of
$R_{\rm N}$, $R_{\rm q}$ and $R$.
The relative variance of avalanche fluctuation ($f\/$) is taken to be 2/3.
The value of $R_{\rm q}$ is almost constant ($\sim 1+f$) for practical pad heights
(\gsim 6 mm) whereas $R_{\rm N}$ is a decreasing function of the pad height as
expected.

\begin{figure}[htbp]
\begin{center}
\hspace{10mm}
\includegraphics*[scale=0.80]{\figdir/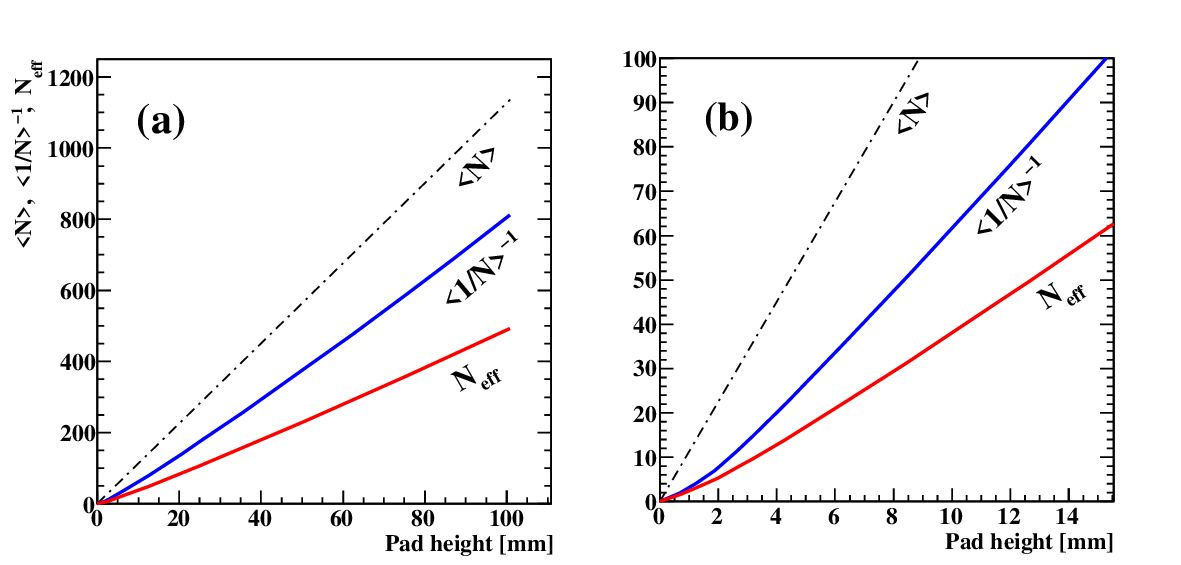}
\end{center}
\caption{\label{figA2}
\footnotesize (a) Average number of electrons ($\left< N \right>$),
inverse of the average of $1/N$ ($\left< 1/N \right>^{-1}$),
and effective number of electrons ($N_{\rm eff}$)
as a function of the pad height.
(b) Details for small pad heights. 
}
\end{figure}

The values of $\left< 1/N \right>^{-1}$ (= $\left< N \right>/R_{\rm N}$)
and $N_{\rm eff}$ are plotted in Fig.~\ref{figA2} against the pad
height, along with $\left < N \right >$.
It is clear that $N_{\rm eff}$ is not a linear function of the pad
height because of $R_{\rm N}$ decreasing with the pad height.
The effective number of electrons is about 31\% (43\%)
of $\left< N \right>$ for a pad height of 6.3 mm (100.8 mm).

\section{Derivation of the function $g$}
In this appendix we derive the explicit expression of the function $g(\sigma/h)$
used to evaluate Var($\nu$) and Var($N$) in Section 4. The function is defined as
\begin{displaymath}
h \cdot g\left( \frac{\sigma}{h} \right) = \lim_{H \to \infty} 
    \int_{-H/2}^{H/2} \Pi^2(Y)\;dY = \int_{-\infty}^\infty \Pi^2(Y)\;dY
\end{displaymath}
where
\begin{displaymath}
\Pi (Y) = \int_{-h/2}^{h/2} G(y;Y,\sigma)\;dy
\end{displaymath}
with
\begin{displaymath}
G(y;Y,\sigma) \equiv \frac{1}{\sqrt{2\pi} \sigma} \cdot 
          \exp \left[ - \frac{(y-Y)^2}{2\sigma^2} \right] \; .
\end{displaymath}

Let us carry out the integration on the right hand side of the equation:
\begin{eqnarray*}
\int_{-\infty}^\infty \Pi^2(Y)\;dY &=&
\frac{1}{2\pi\sigma^2} \int_{-\infty}^\infty
\left[ 
\int_{-h/2}^{h/2} \exp \left( -\frac{(y-Y)^2}{2\sigma^2} \right) dy
\cdot
\int_{-h/2}^{h/2} \exp \left( -\frac{(y^\prime -Y)^2}{2\sigma^2} \right)
dy^\prime
\right] dY \\
&=& 
\frac{1}{2\pi\sigma^2} \int_{-\infty}^\infty dY
\int_{-h/2}^{h/2} dy \int_{-h/2}^{h/2} dy^\prime \;
\exp \left[
-\frac{1}{2\sigma^2} 
\left( 
(y - Y)^2 + (y^\prime - Y)^2
\right) 
\right]  \\
&=&
\frac{1}{2\pi\sigma^2} \int_{-\infty}^\infty dY
\int_{-h/2}^{h/2} dy \int_{-h/2}^{h/2} dy^\prime \;
\exp \left[
-\frac{1}{\sigma^2} \left(
Y - \frac{y+y^\prime}{2}
\right)^2
- \frac{1}{4\sigma^2}(y - y^\prime)^2
\right] \\
&=&
\frac{1}{2\sqrt{\pi}\sigma} 
\int_{-h/2}^{h/2} dy \int_{-h/2}^{h/2} dy^\prime \;
\exp \left[
- \frac{1}{4\sigma^2}(y - y^\prime)^2 
\right] \\
&=&
\frac{2}{\sqrt{\pi}\sigma} 
\int_0^{h/\sqrt{2}} d\xi
\int_0^{h/\sqrt{2}-\xi}
\exp \left(
- \frac{\eta^2}{2\sigma^2}
\right) d\eta \\
& &
\hspace{65mm} {\rm with \,\,\,} \xi = \frac{(y^\prime + y)}{\sqrt{2}}
 \;\;
{\rm and \,\,} \eta =  \frac{(y^\prime - y)}{\sqrt{2}} \nonumber \\
&=&
\frac{2h}{\sqrt{\pi}} \int_0^1 du 
\int_0^{h \cdot (1-u)/2\sigma} \exp(-v^2)\;dv \\
& & \hspace{65mm} {\rm with \,\,\,} u =\frac{\sqrt{2}}{h} \xi \;\;
{\rm and \,\,} v = \frac{\eta}{\sqrt{2}\sigma} \nonumber \\
&=&
h \int_0^1 f(u)\;du \\
& &
\hspace{65mm} {\rm with \,\,\,} f(u) = \frac{2}{\sqrt{\pi}}
\int_0^{h \cdot (1-u)/2\sigma} \exp(-v^2) \, dv \\
&=& h \cdot \left( \Bigl[ u f(u) \Bigr]_0^1
- \int_0^1 u f^\prime (u)\;du
 \right) \\
&=& 
\frac{h^2}{\sqrt{\pi}\sigma}
\int_0^1 u \cdot \exp \left[- \left( \frac{h \cdot (1-u)}{2\sigma}
			      \right)^2  \right] du \\
&=& \frac{2h}{\sqrt{\pi}}
\int_0^{h/2\sigma}\left( 1 - \frac{2\sigma}{h} t \right) \cdot
\exp(-t^2)\;dt \\
& &
\hspace{65mm} {\rm with \,\,\,} t = \frac{h \cdot (1 - u)}{2\sigma} \\
&=&
h \cdot \left( {\rm erf} \left(\frac{h}{2\sigma}\right)
 - \frac{4\sigma}{\sqrt{\pi}h}
\int_0^{h/2\sigma} t \cdot \exp(-t^2)\;dt
\right) \\
&=&
h \cdot \left( {\rm erf} \left(\frac{h}{2\sigma}\right)
 - \frac{2\sigma}{\sqrt{\pi}h}
\int_0^{h^2/4\sigma^2} \exp(-w)\;dw
\right) \\
& &
\hspace{65mm} {\rm with \,\,\,} w = t^2 \\[5mm]
&=& h \cdot \left[
{\rm erf} \left(\frac{h}{2\sigma}\right) + 
\frac{2\sigma}{\sqrt{\pi}h} \cdot
\left(
\exp \left(-\frac{h^2}{4\sigma^2}
\right)
-1 \right)
\right] \; .
\end{eqnarray*}
Hence,
\begin{displaymath}
g \left(\frac{\sigma}{h}\right) = 
{\rm erf} \left(\frac{h}{2\sigma}\right) - 
\frac{2}{\sqrt{\pi}} \cdot \frac{\sigma}{h} \cdot
\left[
1 - \exp \left(-\frac{h^2}{4\sigma^2}\right) 
\right] \; .
\end{displaymath}

It should be noted that
the error function is given by
\begin{displaymath}
{\rm erf}(x) = \frac{2}{\sqrt{\pi}} \int_0^x e^{-t^2} \; dt
\end{displaymath}
and that
\begin{displaymath}
g(0) = {\rm erf}(\infty) = 1 \;\;\; {\rm and} \;\;\;
g(\infty) = 0 \; .
\end{displaymath}


\end{document}